\documentclass[twocolumn]{aastex631}

\usepackage{siunitx}
\usepackage{natbib}
\usepackage{bm}
\usepackage{mathtools}
\usepackage{multirow}
\usepackage{CJK}         
\usepackage{hyperref}

\defcitealias{Kokubo:2024}{KH24}
\defcitealias{Tee:2024}{T24}

\begin{document}
\begin{CJK*}{UTF8}{gbsn}

\title{Do Little Red Dots Vary?}

\author[0000-0002-1174-2873]{Amy Secunda}\thanks{E-mail: asecunda@flatironinstitute.org}
\affil{Center for Computational Astrophysics, Flatiron Institute, New York, NY 10010, USA}

\author[0000-0002-6748-6821]{Rachel S. Somerville}
\affil{Center for Computational Astrophysics, Flatiron Institute, New York, NY 10010, USA}

\author[0000-0002-2624-3399]{Yan-Fei Jiang (姜燕飞)}
\affil{Center for Computational Astrophysics, Flatiron Institute, New York, NY 10010, USA}

\author[0000-0002-5612-3427]{Jenny E. Greene}
\affil{Department of Astrophysical Sciences, Princeton University, Peyton Hall, Princeton, NJ 08544, USA}

\author[0000-0001-6278-032X]{Lukas J. Furtak}
\affiliation{Department of Physics, Ben-Gurion University of the Negev, P.O. Box 653, Be'er-Sheva 84105, Israel}

\author[0000-0002-0350-4488]{Adi Zitrin}
\affiliation{Department of Physics, Ben-Gurion University of the Negev, P.O. Box 653, Be'er-Sheva 84105, Israel}

\begin{abstract}
Little red dots (LRDs), high-redshift, compact, red objects with V-shaped spectra, are one of the most exciting and perplexing discoveries made by the James Webb Space Telescope (JWST). While the simplest explanation for LRDs is that they are high redshift active galactic nuclei (AGN), due to their compactness and frequent association with broad line emission, the lack of corresponding X-ray emission and observed variability cast doubt on this picture. Here, we simulate LRD light curves using both traditional models for sub-Eddington AGN variability derived empirically from lower-redshift AGN observations and moderately super-Eddington AGN disk models from radiation magnetohydrodynamic simulations to examine the reason for the lack of variability. We find that even though most LRDs have only been observed 2--4 times in a given waveband, we should still be detecting significantly more variability if traditional sub-Eddington AGN variability models can be applied to LRDs. Instead, our super-Eddington model light curves are consistent with the lack of observed LRD variability. In addition, the ongoing high-cadence {\sc nexus} campaign will detect changes in magnitude, $\Delta m>1$, for traditional sub-Eddington models, but will only observe significant continuum variability for the lowest mass LRDs for our super-Eddington AGN models. Even if LRDs lack continuum variability, we find that the ongoing spectroscopic JWST campaign {\sc twinkle} should observe broad emission line variability as long as soft X-ray irradiation manages to reach the broad line region from the inner disk. Our models show that super-Eddington accretion can easily explain the lack of continuum variability in LRDs.
\end{abstract}

\section{Introduction}
\label{sec:intro}

One of the most unexpected results from the James Webb Space Telescope (JWST) is the discovery of the so-called Little Red Dots \citep[LRDs,][]{Furtak:2023,Matthee:2024,barro24,greene24,labbe24,kokorev24a,labbe25,kocevski24,akins24}. These LRDs have a distinct V-shaped spectral energy distribution (SED), with a faint blue UV continuum and a steep rise in continuum flux towards longer rest-frame optical wavelengths. The compactness of these sources \citep{akins24,labbe25,Furtak:2023} and the frequently associated Balmer and Paschen emission lines with $\rm{FWHM}>2000$~km~s$^{-1}$ \citep[e.g.][]{kocevski23,greene24,furtak24b,Matthee:2024,wang24a} have led to the conclusion that these LRDs are obscured high redshift active galactic nuclei (AGN).

At $z>4$ LRDs make up 20--50\% of the AGN population, which is a number density that is about 100 times higher than expected if the UV luminosity functions of $z<4$ quasars are extrapolated down to the fainter magnitudes of LRDs \citep{akins24,greene24,kokorev24a,kocevski24,Hviding:2025}. On the other hand, the number density of LRDs photometrically selected using their V-shaped SEDs and a compactness criterion drops dramatically at $z\lesssim4$ \citep{kocevski24,Ma:2025}. Therefore, determining whether LRDs are indeed AGN and, if so, if their properties can be inferred using lower redshift AGN properties, will have a large impact on understanding the formation of the first supermassive black holes (SMBHs) and their connection to galaxies in the first billion years after the Big Bang.

Several oddities in LRD SEDs cast doubt on this AGN picture. First, follow-up X-ray observations have failed to detect significant X-ray emission associated with LRDs \citep[e.g.,][]{ananna24,yue24,kocevski24,Sacchi:2025}. While absorption could potentially account for the lack of detected X-rays, the lack of hot dust observed in the MIR is inconsistent with expectations from known obscured AGN \citep{akins24,williams24,Setton:2025,Casey:2024}. The SEDs of these objects also have continuum inflection at the Balmer limit or occasionally even distinct Balmer breaks that are hard to explain with dust alone \citep[e.g.,][]{furtak24b,labbe24,wang24a,Ma:2025,setton24}. These spectral features, as well as gas absorption signatures in broad emission lines, suggest that LRDs are surrounded by a dense gas absorber \citep{inayoshi24a,ji2025,Naidu:2025,Maiolino:2025}. 

Also puzzling is the lack of variability detected at rest-frame UV and optical wavelengths for the vast majority of LRDs with multiple epochs of observations \citep{Kokubo:2024,Zhang:2024,Tee:2024,Furtak:2025}. Detecting variability in multi-epoch LRD observations would be an unambiguous sign that they are indeed AGN, because AGN continuum and broad lines are well known to vary over a broad range of observable timescales from hours to decades, while stellar populations do not. \citet[][hereafter \citetalias{Tee:2024}]{Tee:2024} studied 30 bright LRDs that were also detected in archival images from the Hubble Space Telescope (HST), providing them with two epochs separated by 6--11 years in two different rest-frame UV wavebands that they color-corrected to match HST and JWST filters. They found on average a change in magnitude of $\Delta m=0.15\pm0.26$~mag, which is not significant relative to the mean magnitude error of the HST observations, $\sim0.19$. \cite{Zhou:2025} simulated the results of \citetalias{Tee:2024} using their corona-heated accretion-disk reprocessing model. They found that the observed magnitude variations are dominated by the HST measurement uncertainties and that the lack of observed rest-frame UV variability could be explained by high AGN luminosity or strong galaxy contamination.

\citet[][hereafter \citetalias{Kokubo:2024}]{Kokubo:2024} looked at 3 LRDs and 2 less compact H$\alpha$ emitters with 2--4 epochs of JWST observations spanning $<2$ years in the observer-frame. They found at most a maximum change in UV or optical magnitude between two observations of $\Delta m\approx 0.1$~mag, and no changes in UV or optical magnitude between different epochs a factor of two greater than their photometric errors. \cite{Zhang:2024} performed the most comprehensive analysis of LRD variability to date. They compiled a sample of 314 LRDs with at least two epochs of observations with JWST spanning $<2$ years in the observer-frame. They found that the observed changes in UV and optical magnitudes between different epochs were consistent on the whole with no variability. They only found eight LRDs with evidence of variability, and only two LRDs where the variability is correlated among multiple wavebands. 

Finally, \cite{Furtak:2025} studied the line, UV, and optical variability of the triply imaged lensed LRD, A2744-QSO1 \citep[see also,][]{ji2025}. Due to the lensing-induced time delays between the images they were able to examine the variability over 6--12 epochs spanning 2.7 years in the rest-frame. While they found evidence for $\Delta m\approx0.4$~mag between the three images, this change in magnitude is not large enough to overcome the uncertainty in the photometry and the inter-image calibration. On the other hand, they did observe statistically significant changes in the equivalent widths of the H$\alpha$ and H$\beta$ emission lines, which may be a sign of emission line variability in this object \citep[see also][]{ji2025}. 

Because most studies of LRD variability can draw on only a few epochs of observation, it is unclear whether variability is rare for LRDs or if the data are too sparse. Ongoing and upcoming JWST surveys, such as {\sc nexus} \citep[JWST Proposal Cycle 3, ID. 5105,][]{Shen:2024}, will provide an opportunity to examine higher cadence and longer baseline observations of LRDs. In addition, the upcoming {\sc twinkle} campaign (JWST Proposal Cycle 4, ID. 7404PI, PI Naidu), will allow us to look for flux variability in broad Balmer emission lines.

If, however, we continue to not detect variability at the levels predicted by empirical models fit to lower redshift AGN, we propose that super-Eddington accretion could possibly explain this lack of variability, as well as other discrepancies between LRDs and lower redshift AGN. For example, super-Eddington accretion can help account for the preponderance of relatively massive SMBHs at these high redshifts by both accelerating their growth and lowering our current mass estimates \citep{kokorev23,furtak24b,wang24a,lambrides24}. 

Super-Eddington accretion could also contribute to the lack of observed X-ray emission from LRDs by {\sc chandra} by decreasing and softening the X-ray emission of LRDs \citep{inayoshi24b,madau24,madau25,pacucci24,lambrides24}, although recent evidence using stacked {\sc chandra} observations suggests that the observed limits on X-ray emission are too low to be explained by super-Eddington accretion alone \citep{Sacchi:2025}. If gas absorption is required to explain these low levels of X-ray emission, super-Eddington accreting disks are known to drive dense gas outflows and expand the disk photosphere \citep[e.g.,][]{Shakura1973,King:2003,Ohsuga:2005,Skadowski:2015,Hu:2022,Jiang:2024,Jiang:2019b,Zhang:2025}. This puffed up disk photosphere could perhaps resemble the proposed ``black hole star'' model, where the AGN is fully embedded in dense gas \citep{inayoshi24a,Naidu:2025,Maiolino:2025,Liu:2025}. 

In this paper, we contextualize previous and future LRD variability studies by predicting the likelihood of observing significant LRD variability given the cadence of observations and two different AGN models for LRDs. The first model is an empirical model based on observations of lower redshift sub-Eddington AGN. The second model is a theoretical model based on radiation magnetohydrodynamic (MHD) simulations of moderately super-Eddington AGN. In Section \ref{sec:methods}, we describe how we generate mock LRD light curves using both this empirical (Section \ref{sec:methods:empirical}) and theoretical (Section \ref{sec:methods:theoretical}) approach. We also describe in Section \ref{sec:methods:emission} how we use both of these approaches to generate mock broad emission line light curves to examine whether we will be able to detect emission line variability with {\sc twinkle}. In Section \ref{sec:results} we show the predictions of our two models for the likelihood of observing significant variability with current (Section \ref{sec:results:current}) and ongoing (Section \ref{sec:results:future}) observations. Finally, we discuss and summarize our results in Sections \ref{sec:discuss} and \ref{sec:conclude}, respectively.

\section{Methods}
\label{sec:methods}

Detecting or not detecting variability for LRDs could be a powerful way to constrain different LRD models. Here we study two different models for LRDs. The first model assumes that LRDs are well represented by empirical models for lower redshift AGN. We provide details on how we generate mock LRD light curves using this empirical model in Section \ref{sec:methods:empirical}. The second model assumes that LRDs are super-Eddington accreting AGN. In this second case, LRD variability cannot be modeled using standard empirical models, and we develop a model for LRD variability using radiation MHD simulations of super-Eddington AGN. We describe this model in detail and how we use it to generate mock LRD light curves in Section \ref{sec:methods:theoretical}. In Section \ref{sec:methods:emission} we describe how we use both of these models to simulate mock broad Balmer line emission light curves, which we will use to study how these underlying models impact our ability to observe variability in broad line emission.

\subsection{Empirical Damped Random Walk Model}
\label{sec:methods:empirical}

In this section, we assume that LRD variability is well modeled by empirical models of variability fit to lower redshift AGN. Rest-frame UV and optical continuum variability have been studied extensively for lower redshift AGN with masses ranging from $10^5-10^{10}~M_{\odot}$ and primarily sub-Eddington luminosities of around $0.02 - 0.5~L_{\rm edd}$. This continuum variability is most frequently modeled as a Damped Random Walk (DRW) \citep[e.g.,][]{Kelly:2009,MacLeod:2010,Burke2021,Burke:2023,Stone:2022}. A DRW is a stochastic process defined by a power spectral density (PSD) that scales with the frequency of the variability as PSD~$\propto\nu^{-2}$ at high frequency, and transitions to white noise at frequencies below a characteristic damping timescale, $\tau$. The structure function at infinity, SF$_{\infty}$, parameterizes the magnitude of the variability at infinitely long timescales.

We calculate $\rm{SF}_{\infty}$ using the parameterization as a function of rest-frame wavelength, $\lambda_{\rm RF}$, \emph{i}-band absolute magnitude, $M_{\rm i}$, and SMBH mass, $M_{\rm BH}$ from \cite{Burke:2023}, 
\begin{equation}
\label{eq:sfinf}
\begin{split}
    \log\left(\frac{\rm{SF}_{\infty}}{\rm{mag}}\right) = A + B\log\left(\frac{\lambda_{\rm RF}}{4000~\dot{\rm{A}}}\right) + C(M_i+23) + \\
    D\left(\frac{M_{\rm BH}}{10^9~\rm{M}_{\odot}}\right),
    \end{split}
\end{equation}
where $A=-0.51\pm0.02$, $B=-0.479\pm0.005$, $C=0.131\pm0.008$, and $D=0.18\pm0.03$. To calculate $M_{\rm i}$ we assume a typical Eddington ratio of $L/L_{\rm edd}=0.1$, because the bolometric luminosities of LRDs are still uncertain. We determine the damping timescale using the mass-dependent parameterization from \cite{Burke2021},
\begin{equation}
\label{eq:tau}
    \tau = 107\left( \frac{M_{\rm BH}}{10^8~\rm{M}_{\odot}} \right)^{0.38}.
\end{equation}

We use {\sc EzTao} \citep{Yu:2022} to generate mock DRW light curves with parameters corresponding to various masses, redshifts, and rest-frame wavelengths for observed LRDs. {\sc EzTao} is a Python toolkit that uses {\sc celerite}, a fast Gaussian process regression library, to generate different representations of AGN variability, including the DRW. We generate these DRWs with high even cadences and then subsample these cadences to match the reported observation times for different LRDs in \citetalias{Kokubo:2024} and \citetalias{Tee:2024} and the anticipated cadences for {\sc nexus}. 

For our mock light curves based on the LRDs in \citetalias{Kokubo:2024} we simulate 1000 observer-frame F115W-band and 1000 observer-frame F356W-band light curves using the reported masses, redshifts, and observation times for four of the \citetalias{Kokubo:2024} LRDs.\footnote{We exclude MSAID2008 because it is no longer thought to be a LRD.} For mock \citetalias{Tee:2024} light curves we make 1000 mock observer-frame F115W-band light curves with masses randomly drawn from a log-uniform distribution ranging from $M_{BH}=10^5$ to $10^8$~M$_{\odot}$. We then mock observe these light curves at two epochs using the mean observing time for the JWST observation and the mean observing time for the EGS, GOODS-S, or Abell 2744 HST fields given in \citetalias{Tee:2024} randomly scattered by $\pm 20$~days. For our mock {\sc nexus} light curves we simulate 1000 observer-frame F115W-band and 1000 observer-frame F356W-band light curves with masses randomly drawn from a log-uniform distribution ranging from $M_{BH}=10^5$ to $10^8$~M$_{\odot}$. We mock observe these light curves with the anticipated bi-monthly cadence randomly scattered by $\pm 14$~days. 

We simulate our mock \citetalias{Tee:2024} and {\sc nexus} light curves at redshift, $z=6$, because it puts the F115W-band and the F356W-band in the rest-frame UV and optical, respectively, and is a good representation for the redshift of many LRDs. Decreasing the redshift will increase both the rest-frame observation baseline and wavelength. Testing different redshifts $4\leq z \leq 9$, we find that these two factors tend to cancel out, so varying the redshift has little impact on the observed variability. The amount of optical variability could be maximized by looking at the lowest redshift LRDs in the shortest wavelength bands, but for now we restrict our examination to one rest-frame optical and one rest-frame UV band. Because it has a larger impact on the amount of variability we expect to detect, SMBH mass is the main parameter we vary for our DRW model light curves. We show an example DRW model optical light curve for a $10^8~M_{\odot}$ AGN at $z=6$ in the top panel of Figure \ref{fig:exlcs} in Appendix \ref{appendix}.

\subsection{Super-Eddington Model}
\label{sec:methods:theoretical}

\begin{figure*}
    \centering
    \includegraphics[width=\columnwidth]{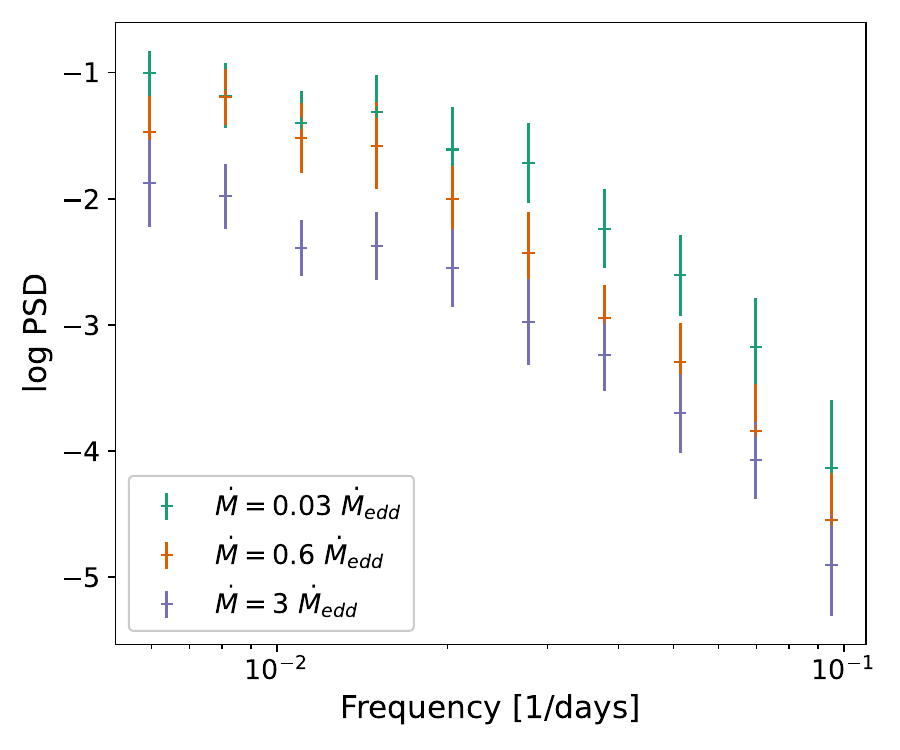}
    \includegraphics[width=\columnwidth]{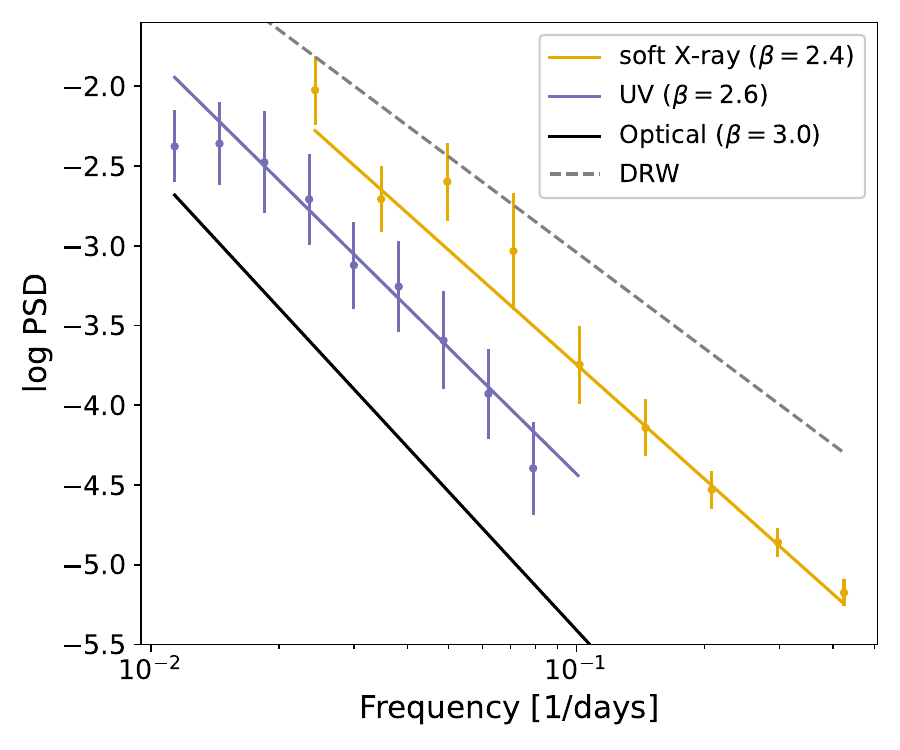}
    \caption{\textbf{Left panel:} The frequency-binned variability PSD of the UV light curves emitted from $<200~r_{\rm g}$ for three different {\sc athena++} simulations from \cite{Jiang:2025} labeled by their mass accretion rate. \textbf{Right panel:} The frequency-binned variability PSD of the soft X-ray (in yellow, emitted from $\sim 10~r_{\rm g}$) and UV (in purple, emitted from $<200~r_{\rm g}$) light curves from the simulations in \cite{Jiang:2019b} and \cite{Jiang:2025}, respectively, with $\dot{M}=3~\dot{M}_{\rm edd}$. We also show our best fit power-law for each PSD, with best-fit parameter in the label. The black solid line shows the extrapolated PSD for the optical emitting region at $10^4~r_{\rm g}$. The dashed black line shows the optical DRW PSD for the same mass SMBH ($M_{BH}=10^8~M_{\odot}$), at 10\% Eddington. For both panels the error bars represent the standard deviation of that frequency bin. The simulations show that higher accretion rates lead to lower variability, especially compared to a sub-Eddington DRW model.}
    \label{fig:super_psd}
\end{figure*}

Several of the more puzzling features of LRDs could be more easily explained if they are super-Eddington accretors. In particular, super-Eddington accretion relaxes the need for very heavy black hole seeds \citep[e.g.][]{furtak24b,greene24} and can help account for low levels of X-ray emission \citep{inayoshi24b,madau24,madau25,pacucci24,lambrides24} and spectral features that suggest significant gas absorption \citep[e.g.,][]{Naidu:2025,Maiolino:2025,Liu:2025,Kido:2025}, because super-Eddington accretion expands the disk photosphere and leads to optically thick outflows and intrinsically weak high-energy flux. Theoretical models also suggest that super-Eddington accreting AGN may be less relatively variable at UV and optical wavelengths \citep{inayoshi24b}. In this section, we outline how we use radiation MHD simulations of AGN disks from \cite{Jiang:2019b} and \cite{Jiang:2025} to create a model for the variability of super-Eddington LRDs and generate mock LRD light curves.

\subsubsection{Radiation MHD Simulations}

\cite{Jiang:2025} use {\sc athena++} and the radiation scheme in \cite{Jiang:2021} to perform global three-dimensional radiation MHD simulations of the UV emitting region (50--200~$r_{\rm g}$) of $M_{BH}=10^8~M_{\odot}$ AGN disks. Radiation is generated based on the local temperature of the disk and varies as a function of time due to the internal physics of the disk, such as magnetorotational instability (MRI) turbulence, convection, or a magnetocentrifugal wind. Note, that these instabilities are generated self-consistently by evolving the equations of ideal MHD and therefore the variability is not directly set by any empirical or theoretical model. Light curves are made by integrating the outgoing flux from the inner 50--200~$r_{\rm g}$ of the simulated AGN disk. We de-trend these simulated light curves by fitting and then subtracting off a best-fit power-law to remove long term trends resulting from the simulation set up. 

There is no irradiation from a central X-ray emitter/corona in these simulations. However, super-Eddington accretors are expected to be X-ray weak and have expanded disk photospheres. These expectations are in good agreement with the lack of detected X-rays for a majority of LRDs \citep{ananna24,yue24,kocevski24}, which has led to the conclusion that LRDs are either embedded in dense gas \citep[e.g.,][]{pacucci24,inayoshi24a,madau25,Naidu:2025} or intrinsically X-ray weak \citep[e.g.,][]{inayoshi24b}. Either way, the amount of X-ray flux reaching the UV and optical emitting region of the disk will be low. \cite{Secunda:2025} showed that insufficient X-ray irradiation prevents X-rays from driving variability in UV-optical light curves emitted from AGN disks. Therefore, in our super-Eddington model we make the assumption that the observed optical variability from LRDs will be local intrinsic variability like the variability in these \cite{Jiang:2025} simulations.

We show the PSDs of light curves from three different simulations in \cite{Jiang:2025} in the left panel of Figure \ref{fig:super_psd}. After evolving to a steady-state, each simulation has a different mass accretion rate ranging from $\dot{M}=0.03~\dot{M}_{edd}$ to $\dot{M}=3~\dot{M}_{edd}$, due to different initial conditions for the density and magnetic field parameters. The simulation with $\dot{M}=0.03~\dot{M}_{edd}$ has the highest variability at most frequencies, while the simulation with $\dot{M}=3~\dot{M}_{edd}$ has the lowest variability. In other words, accretion rate and variability are anti-correlated. 

This anti-correlation occurs because in these simulations from \cite{Jiang:2025} the total optical depth across the disk is larger when the mass accretion rate is increased. In addition, an increase in the accretion rate pushes the photosphere where the radiation is emitted out to farther radii. At larger radii the dynamical and thermal timescales that set the characteristic variability timescales will be longer. As a result, more of the variability ends up on timescales that are longer than probed by these simulations and current LRD observations. An extended photosphere is a common feature of super-Eddington AGN radiation MHD simulations \citep[e.g.][]{Skadowski:2015,Jiang:2019b,Zhang:2025}.

\subsubsection{Optical Light Curves}

In this paper, we focus on the optical variability of super-Eddington LRDs. In our super-Eddington model we expect UV emission to be either absorbed or scattered depending on the absorption and scattering optical depth of the extended photosphere. If the emission is absorbed, as could be the case for LRDs with strong Balmer breaks, we do not expect to observe any UV variability. If instead the UV is scattered, the short timescale variability will likely be damped up to the characteristic timescale near the photosphere. Future work should examine what this variability might look like. 

The simulations from \cite{Jiang:2025} are of the UV emitting region of the AGN disk. Fully resolved radiation MHD simulations of the optical emitting region of an AGN disk are computationally challenging because of the large radial extent of the optical region. However, it is generally expected that the typical variability timescales of photons of different wavelengths should be directly related to the distance where they are produced. Therefore, in order to simulate mock optical LRD light curves, we extrapolate existing simulations of the soft X-ray and UV emitting disk to the optical emitting disk.

In the right panel of Figure \ref{fig:super_psd} we compare the PSD of the simulation from \cite{Jiang:2025} with $\dot{M}=3~\dot{M}_{edd}$ (in purple) to a simulation from \cite{Jiang:2019b} with the same accretion rate (in yellow). These two simulations are similar, except the simulation in \cite{Jiang:2019b} is of the inner ($<10~r_g$) soft X-ray emitting region of the disk. We fit a simple power-law, PSD~$\propto C_0 \nu^{-\beta}$, to the light curves over the overlapping frequency range of the simulations. The simulation of the soft X-ray emitting region of the disk has a shallower PSD, $\beta=2.4$, than the simulation of the UV emitting region, $\beta=2.6$. This result is anticipated because the dynamical and thermal timescales will be shorter in the inner disk allowing for more high frequency variability. We also find $C_0=10^{-6}$ and $C_0=10^{-7}$ for the soft X-ray and UV emitting regions, respectively, which means overall the simulation of the soft X-ray emitting region has more variability than the simulation of the UV emitting region.

In our super-Eddington model, the optical emitting region will be at $\sim 10^4 - 10^5~r_{\rm g}$ \citep{Liu:2025}. We logarithmically extrapolate the values for $\beta$ and the structure function we fit to the soft X-ray and UV light curves and find $\beta=3.0$ and $C_0=5\times10^{-9}$ for optical light curves emitted at $10^4~r_{\rm g}$. We show this optical PSD as the solid black line in the right panel of Figure \ref{fig:super_psd}. The PSD for this super-Eddington model has significantly less variability over day to month timescales than an empirically modeled optical DRW PSD for a $10^8~M_{\odot}$ AGN at 10\% Eddington, which we show for comparison as a dashed line in the right panel of Figure \ref{fig:super_psd}. The variability is lower not just because the higher accretion rate lowers the magnitude of the variability, but also because the super-Eddington PSD only includes locally generated variability while the DRW model assumes that the X-rays are driving the variability of the optical light curve.

Because we do not expect the variability to continue as a power-law to the lowest frequencies, we need to estimate a damping timescale at which we can expect the PSD to flatten, becoming shallower than $\beta=3.0$. If we fit a bending power-law to the UV PSD, we find evidence of a damping timescale around 100~days. The damping timescale should scale with the local dynamical timescales, $\propto r^{3/2}$, giving a damping timescale of around $4\times10^4$~days in the optical emitting region.  Our derived damping timescale is consistent with the expected thermal timescale in this region of the disk. In order to cap the long timescale variability, we therefore set the length of our simulated light curves to this damping timescale of $4\times10^4$~days before sub-sampling them with current and anticipated LRD observing cadences.

We use our extrapolated values for the super-Eddington PSD to generate 1000 mock rest-frame optical, observer-frame F356W-band, light curves using the method in \cite{Timmer1995}. We mock observe these light curves with the observing cadences of four example LRDs from \citetalias{Kokubo:2024} and the planned {\sc nexus} cadence. As in the previous section, we use the given redshifts in \citetalias{Kokubo:2024} for the four example LRDs. For our mock {\sc nexus} light curves we chose a redshift $4<z<9$ from a random uniform distribution. While varying the redshift for these mock super-Eddington light curves broadens the distributions slightly, ultimately the redshift does not have a large impact on the amount of mock observed variability. We show an example mock super-Eddington optical light curve for a $10^8~M_{\odot}$ AGN at $z=6$ in the bottom panel of Figure \ref{fig:exlcs} in Appendix \ref{appendix}.

\begin{figure*}
    \centering
    \includegraphics[width=\linewidth]{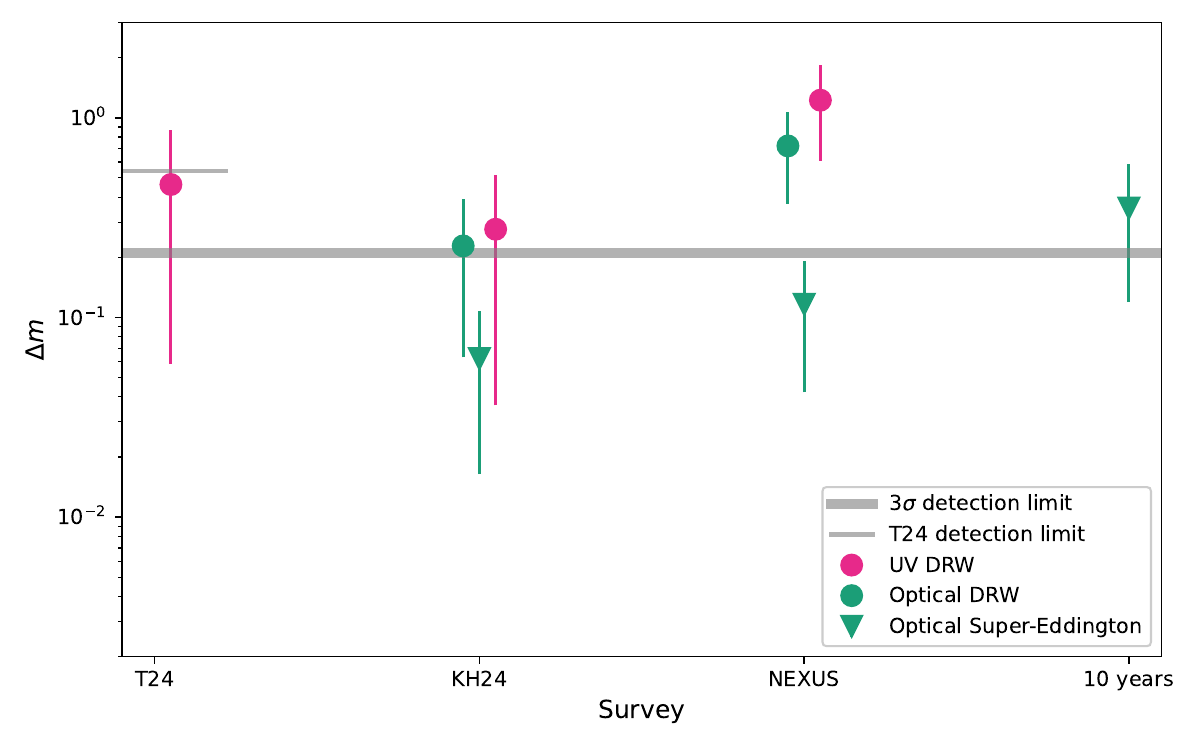}
    \caption{The mean maximum difference in magnitude, $\Delta m = \max(m) - \min(m)$, between all epochs for 1000 simulated observer-frame F115-band (F356-band) light curves in pink (green) mock observed with different previous and ongoing LRD observation cadences. Closed circles represent $\langle \Delta m \rangle$ for our sub-Eddington DRW model light curves and triangles represent $\langle \Delta m \rangle$ for our super-Eddington model light curves. Error bars show the standard deviations of $\Delta m$ for the 1000 mock light curves. The thick gray line shows a $\sim 3\sigma$ detection limit for variability based on the current precision of available JWST photometry, while the thin gray line shows this limit for LRDs in \citetalias{Tee:2024} which is higher due to the lower depth of HST observations. If LRD variability can be modeled using DRW models fit to lower redshift AGN, we should have already observed variability in the rest-frame UV and optical for more LRDs, and we will definitely observe variability for LRDs with {\sc nexus}. On the other hand, if they are super-Eddington accretors, we do not expect to detect variability for most LRDs in either current or ongoing observations, until we can reach a baseline of up to 10~years.}
    \label{fig:summary}
\end{figure*}

\subsection{Mock Broad Line Emission}
\label{sec:methods:emission}

When ionizing X-ray and UV emission irradiates the broad line region (BLR), the variability in these continuum light curves is echoed in the broad emission lines producing broad line variability \citep{Blandford:1982,Peterson2004}. To make predictions for our future ability to observe this broad line variability for LRDs, we simulate mock broad Balmer emission line light curves with a cadence similar to the upcoming JWST NIRCAM campaign, {\sc twinkle}. {\sc twinkle} will observe 9 LRDs three times over Cycle 4 and compare line fluxes between these three observations and an additional Cycle 1 observation. Total, there will be four observations over roughly a rest-frame year.  To simulate broad line variability, we generate a driving mock X-ray or UV light curve, reprocess this light curve with a Gaussian response function to get the emission line variability, and finally mock observe it with the proposed {\sc twinkle} cadence. We use three different models for the driving X-ray or UV light curves. 

The first model is an empirical model based on lower redshift sub-Eddington AGN, where the X-ray light curve is the driving light curve. This model assumes that, while we do not observe X-ray emission from LRDs, X-ray emission is still able to reach the BLR. X-ray variability is commonly modeled with a bending power-law PSD \citep[e.g.,][]{Papadakis:2002,McHardy:2004,McHardy2007,Gonzalez:2012,Georgakakis:2021,Jaiswal:2023}. Fits to X-ray observations give a high frequency slope of $-2$ and a low frequency slope of $-1$ around the bending frequency $\nu_b=0.29\left(\frac{M_{BH}}{10^8~M_{\odot}}\right)^{-1}$~days$^{-1}$ and a constant amplitude $A=0.02$ \citep{Paolillo:2023}. We generate 1000 X-ray driving light curves with SMBH masses drawn from a log-uniform distribution from $M_{BH}=10^5$ to $10^8~M_{\odot}$ at $z=5.3$. We use $z=5.3$ for these mock light curves because that is the average redshift of the 9 LRDs that will be monitored by {\sc twinkle}, which are all roughly between $5<z<5.5$. 

We also generate $z=5.3$ driving light curves using our best-fit models in the previous section (see right panel of Figure \ref{fig:super_psd}) for the soft X-ray and UV light curves from the super-Eddington radiation MHD simulations in \cite{Jiang:2019b} and \cite{Jiang:2025}, respectively. We use both the soft X-ray and UV light curves from these simulations, because in super-Eddington AGN it is very possible that the soft X-ray will be absorbed before reaching the BLR. Using both wavelength light curves allows us to test which could be responsible for driving line variability in a super-Eddington LRD. Understanding the driver behind emission line variability is important because it could allow us to probe emission regions of LRDs we are unable to observe directly. 

 We reprocess our driving X-ray, soft X-ray, and UV light curves with a Gaussian response function to simulate the broad line response light curve. There is no simple empirical model for BLR reprocessing, but a Gaussian response function is commonly used to model the reprocessing of the driving light curve into the broad line response light curve \citep[e.g.,][]{Czerny:2023,Jaiswal:2023}. The most important parameter of this response function that will impact our ability to detect broad line variability is the Gaussian width, which will set the smoothing of variability on shorter timescales. This smoothing could impact our ability to detect variability with the limited {\sc twinkle} cadence and baseline. SDSS AGN suggest a typical reprocessing width is $\lesssim 100$~days for H$\beta$ lines \citep{Sun:2015}, while observations of individual AGN suggest widths of anywhere from around 5 to 100 days for H$\alpha$ and H$\beta$ \citep{Grier:2013,Xiao:2018,Du:2018,Horne:2021}. Due to this uncertainty, for each mock light curve we choose to use a width between 5 and 100~days drawn from a uniform random distribution. We show an example broad emission line light curve for each underlying driving light curve for a $10^8~M_{\odot}$, $z=5.3$ AGN in Figure \ref{fig:exblr} in Appendix \ref{appendix}.

\section{Results}
\label{sec:results}

\subsection{Predictions for Existing Variability Observations}
\label{sec:results:current}

The vast majority of LRDs that currently have multiple epochs of observations do not show signs of significant continuum variability \citep[\citetalias{Kokubo:2024}, \citetalias{Tee:2024},][]{Zhang:2024,Furtak:2025}. However, most of these LRDs have only been observed 2--4 times over less than a year in the rest-frame. As a result, it is not yet clear if the lack of observed variability is due to the poor time-sampling of observations or the need for a new model for LRDs. In this section, we predict the amount of variability we should have already observed given current observations for two different models for LRDs, an empirical sub-Eddington DRW AGN model based on local AGN and a theoretical super-Eddington AGN model based on radiation MHD simulations. 

\subsubsection{UV Variability}
\label{sec:results:current:uv}

We first use our empirical DRW model AGN light curves to predict the UV variability that we expect to have already observed for two LRD studies, \citetalias{Tee:2024} and \citetalias{Kokubo:2024}. We do not model UV variability using our super-Eddington model because we assume UV variability will be damped away by the expanded gas photosphere either through absorption or scattering. In Figure \ref{fig:summary} we show the mean and standard deviation of the maximum change in magnitude, $\Delta m = \max(m) - \min(m)$, we mock observe for 1000 DRW model mock UV \citetalias{Tee:2024} and \citetalias{Kokubo:2024} light curves.

\citetalias{Tee:2024} combined HST and JWST observations to study the change in magnitude in two different color-corrected wavebands over two epochs separated by 6--11 years in the observer-frame ($\sim 8-26$~months in the rest-frame) for 30 LRDs. They found the mean change in magnitude between the two observations, $\Delta m = 0.15 \pm 0.26$~mag, was not significant relative to the mean magnitude error of the HST observations, $\sim 0.19$. The mean change in magnitude we find for our mock light curves, $\langle\Delta m \rangle= 0.46\pm0.4$~mag, is more than triple the mean change in magnitude \citetalias{Tee:2024} found for their observed LRDs. Furthermore, the majority of changes in magnitude \citetalias{Tee:2024} found for their LRDs are $<0.1$, while we find that the majority of changes in magnitude are $>0.1$ for our mock light curves.

The LRDs in \citetalias{Kokubo:2024} have 2--3 observations each in the F115W-band over the course of $\sim2$ months in the rest-frame. We find $\langle\Delta m \rangle= 0.28\pm0.2$~mag for the mock F115W-band light curves we simulate using the masses, redshifts, and observation times of four of the LRDs studied by \citetalias{Kokubo:2024}. This value is approximately an order of magnitude higher than the variability \citetalias{Kokubo:2024} observed for these LRDs. 

The thick gray line in Figure \ref{fig:summary} shows the $3\sigma$ limit we consider a significant variability detection with JWST, or $\Delta m=0.21$~mag. We derive this value using the minimum error value \cite{Zhang:2024} used for $\Delta m$ in their examination of over 300 LRDs with multi-epoch observations from JWST. The thin gray line shows this $3\sigma$ limit for the LRDs in \citetalias{Tee:2024} which is higher due to larger errors for the HST observations. Our DRW model light curves suggest that we should have observed $>3\sigma$ variability for 33\% and 47\% of LRDs in \citetalias{Tee:2024} and \citetalias{Kokubo:2024}, respectively, while both studies do not observe significant variability for any LRDs. These results imply that we should already be detecting UV variability in a greater fraction of LRDs if UV LRD variability can be well-modeled by empirical DRW models fit to lower redshift AGN.

\subsubsection{Optical Variability}
\label{sec:results:current:optical}

There is growing evidence that UV emission from LRDs is dominated by starlight or scattered light from a gas-embedded AGN \citep[\citetalias{Tee:2024};][]{Matthee:2024,greene24,Naidu:2025,Chen:2025,Torralba:2025,Zhou:2025}. Either would make it difficult to detect variability from the AGN disk at UV wavelengths. Therefore, we now turn to the potential to detect rest-frame optical variability. In Figure \ref{fig:summary} we show the average and standard deviation of the maximum change in magnitude, $\Delta m$, for our simulated rest-frame optical, observer-frame F356W-band, light curves for the LRDs studied in \citetalias{Kokubo:2024}. These LRDs have been observed 3--4 times over a few months in the rest-frame. 

For our DRW model light curves we find $\langle\Delta m \rangle= 0.23\pm0.2$~mag, on average an order of magnitude larger than the changes in magnitude observed for these LRDs. In addition, none of the LRDs in \citetalias{Kokubo:2024} showed a $>3\sigma$ change in magnitude, while we are able to detect $>3\sigma$ optical variability for 42\% of our mock light curves. These results are in good agreement with our results for the UV DRW model light curves. Both suggest that the low cadence of current observations alone is not preventing us from observing variability. The lack of $>3\sigma$ optical variability for nearly all LRDs that currently have multi-epoch optical observations already suggests that standard DRW models are not a good fit for LRD variability.

$\Delta m$ is significantly lower for our super-Eddington model light curves. We find $\langle\Delta m \rangle = 0.062 \pm 0.05$~mag, which is well below our $3\sigma$ limit. In fact we find that less than 1\% of our simulated \citetalias{Kokubo:2024} super-Eddington light curves have $\Delta m>3\sigma$. The small changes in magnitude we predict for our super-Eddington model are consistent with the lack of observed significant optical variability for most LRDs \citep[\citetalias{Kokubo:2024};][]{Zhang:2024}. Therefore, our super-Eddington model appears to be a better fit to current observations.

\subsection{Future Variability Observations}
\label{sec:results:future}

Two ongoing JWST campaigns have the potential to greatly improve our ability to detect variability in LRDs. First, {\sc nexus} will observe dozens of LRDs roughly every two months over the course of three years. These 18 observations will provide a much higher cadence and longer baseline than we currently have for most LRDs. The second ongoing JWST campaign is the NIRCam campaign {\sc twinkle}. {\sc twinkle} will acquire spectra for a total of 4 observations over roughly a year in the rest-frame for 9 LRDs, allowing us to study broad emission line variability. In this section, we make predictions for the variability these campaigns will observe for our sub- and super-Eddington models. We show that these two campaigns will help us to constrain the Eddington ratios of LRDs and reveal information on the main source of irradiation and driver of variability in the BLR.

\subsubsection{Continuum Variability}
\label{sec:results:future:continuum}

First, we make predictions for whether the {\sc nexus} survey will observe variability in rest-frame UV and optical continuum light curves if this variability can be modeled using an empirical sub-Eddington DRW model based on lower redshift AGN. We show the mean and standard deviation of the maximum changes in magnitude for our 1000 DRW model mock F115W-band and F356W-band {\sc nexus} light curves in Figure \ref{fig:summary}. The higher cadence and longer baseline of {\sc nexus} more than triples the amount of variability we should expect to observe for LRDs for these DRW models. We find $\langle\Delta m \rangle = 1.2 \pm 0.6$~mag and $\langle\Delta m \rangle = 0.72 \pm 0.4$~mag, for our mock rest-frame UV and optical light curves, respectively. The horizontal gray line shows the $3\sigma$ limit we anticipate for detecting significant variability in the {\sc nexus} survey, or $\Delta m=0.21$~mag. We find that the high {\sc nexus} cadence allows us to detect $\Delta m>0.21$~mag 100\% of the time for rest-frame UV light curves and 97\% of the time for rest-frame optical light curves. 

Figure \ref{fig:summary} also shows our predictions for the mean and standard deviation of $\Delta m$ for our 1000 super-Eddington model, rest-frame optical, {\sc nexus} light curves. The longer observation baseline and higher cadence planned for {\sc nexus} again leads to an increase in $\Delta m$, but $\langle\Delta m \rangle = 0.12 \pm 0.07$~mag is still below the $3\sigma$ limit. In addition, only 12\% of our mock super-Eddington light curves have $\Delta m >3\sigma$. Therefore, if we still do not observe significant variability for a majority of LRDs after the completion of the {\sc nexus} survey, it could be because LRDs are super-Eddington AGN.

\subsubsection{Broad Line Variability}
\label{sec:results:future:line}

\begin{figure}
    \centering
    \includegraphics[width=\columnwidth]{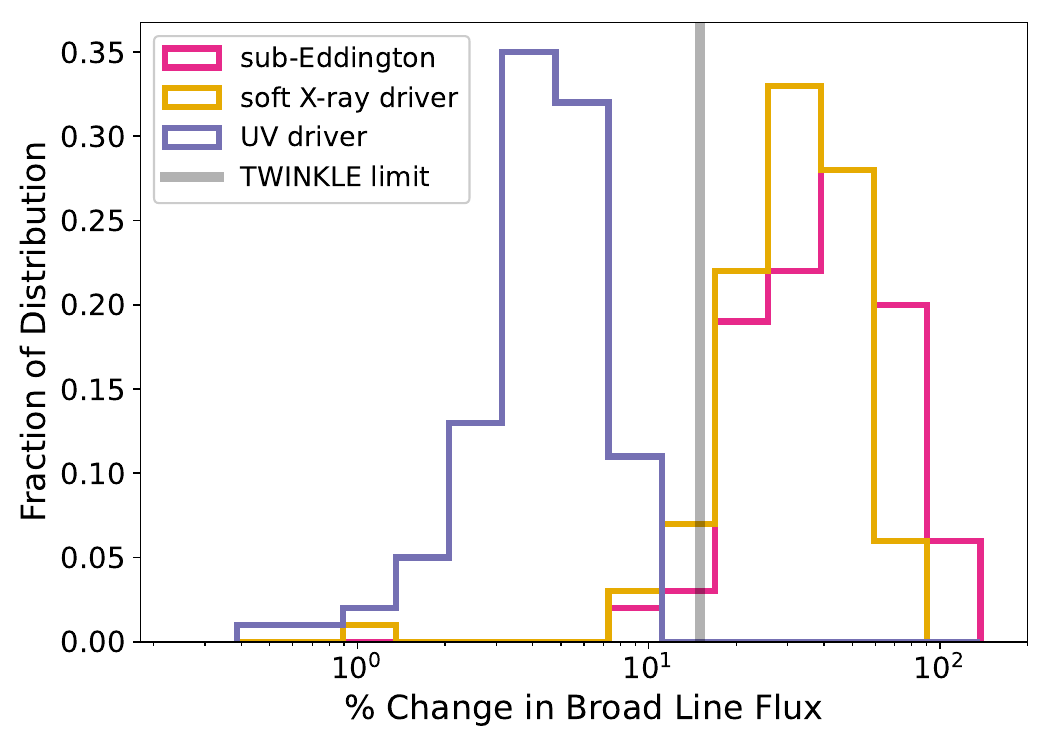}    
    \caption{The distribution of the percent change in broad line flux for 300 mock {\sc twinkle} light curves made assuming three different driving continuum light curves. The pink distribution is for an X-ray driving light curve simulated using empirical models for lower redshift sub-Eddington AGN. The yellow and purple distributions are for soft X-ray and UV driving light curves, respectively, simulated using our best-fit models for the light curves in the radiation MHD super-Eddington AGN disk simulations in \cite{Jiang:2019b} and \cite{Jiang:2025}, respectively. The vertical line shows the $3\sigma$ detection limit for variability in the {\sc twinkle} survey. If LRDs are super-Eddington AGN the driving continuum light curve must be soft X-ray in order to observe broad line variability with the {\sc twinkle} survey.}
    \label{fig:model_twinkle}
\end{figure}

The mock UV and optical LRD light curves from the previous sections indicate that we should already have observed variability for a significant fraction of LRDs if they have variability similar to low redshift AGN. However, there is growing evidence from SED fits, gas absorption signatures in emission lines, and the lack of observed UV variability, that UV emission from these LRDs is dominated by the emission from starlight or at best scattered light from a gas-shrouded AGN \citep[\citetalias{Tee:2024},][]{Matthee:2024,greene24,Naidu:2025,Chen:2025,Torralba:2025,Zhou:2025}. In addition, in the previous sections we showed that if LRDs are super-Eddington accretors, it will be difficult to observe significant optical variability even with higher cadence observing campaigns such as {\sc nexus}. As a result, broad line variability may be the most promising source of observable variability. The detection of broad lines for numerous LRDs indicates that ionizing emission must reach these BLRs, and observed changes in the equivalent widths of broad emission lines already provide evidence of broad line variability in multiple LRDs \citep{Furtak:2025,Naidu:2025,ji2025}. 

Our search for broad line variability will be greatly aided by the ongoing {\sc twinkle} campaign, which will acquire a total of 4 observations over the course of a rest-frame year for 9 LRDs.  When ionizing X-ray and UV continuum emission irradiates the BLR, the variability in this continuum emission drives variability in the broad emission lines. In this section, we compare our expectations for observing broad line variability with {\sc twinkle} given three different driving continuum light curves: an X-ray driving light curve generated using an empirical model based on sub-Eddington lower redshift AGN and a soft X-ray and UV driving light curve from our super-Eddington radiation MHD simulations.

We show the maximum percent change in broad line flux for 1000 mock {\sc twinkle} light curves generated using an empirical model based on lower redshift AGN in pink in Figure \ref{fig:model_twinkle}. {\sc twinkle} expects 5\% line flux errors and should therefore be able to detect any variations in flux greater than 15\%, which we show as the vertical gray line in Figure \ref{fig:model_twinkle}. 96\% of our empirical model mock light curves have a maximum flux variation greater than 15\%. If we randomly select 9 of our sub-Eddington mock AGN light curves for $10^4$ realizations we observe a maximum change in flux $>15\%$ for at least five LRDs 100\% of the time, and for all nine LRDs 70\% of the time.

Figure \ref{fig:model_twinkle} also shows the maximum percent change in flux we expect to observe with {\sc twinkle} if either soft X-ray or UV variability from our super-Eddington radiation MHD simulations is driving broad line variability in yellow and purple, respectively. Our super-Eddington simulations again produce less variability than our models based on sub-Eddington accretors at lower redshifts. Unsurprisingly, the more highly varying soft X-ray irradiation would be able to drive more line variability than the less varying UV radiation.  

In our super-Eddington model, if we randomly select 9 of our mock AGN light curves, for $10^4$ realizations we observe a maximum change in flux $>15\%$ for at least four LRDs 100\% of the time, and for eight or nine LRDs 81\% of the time, if soft X-ray variability is the main driver of broad line variability. If instead UV variability is the main driver, the mean percent change in flux will be roughly an order of magnitude smaller and we would not expect to see any $>15\%$ flux variations at all. Even if we do not reprocess our UV light curves with a response function, we still do not observe any $>15\%$ flux variations because the variability in the UV simulations is too low. Given there is already evidence for broad line variability in LRDs, our results suggest that if LRDs are super-Eddington accretors, then soft X-ray emission from the inner disk must be the main source of irradiation and driver of variability in the BLR. Therefore, broad line variability implies there is soft X-ray emission from LRDs that we are unable to directly observe.

\section{Discussion}
\label{sec:discuss}

\begin{figure}
    \centering
    \includegraphics[width=\columnwidth]{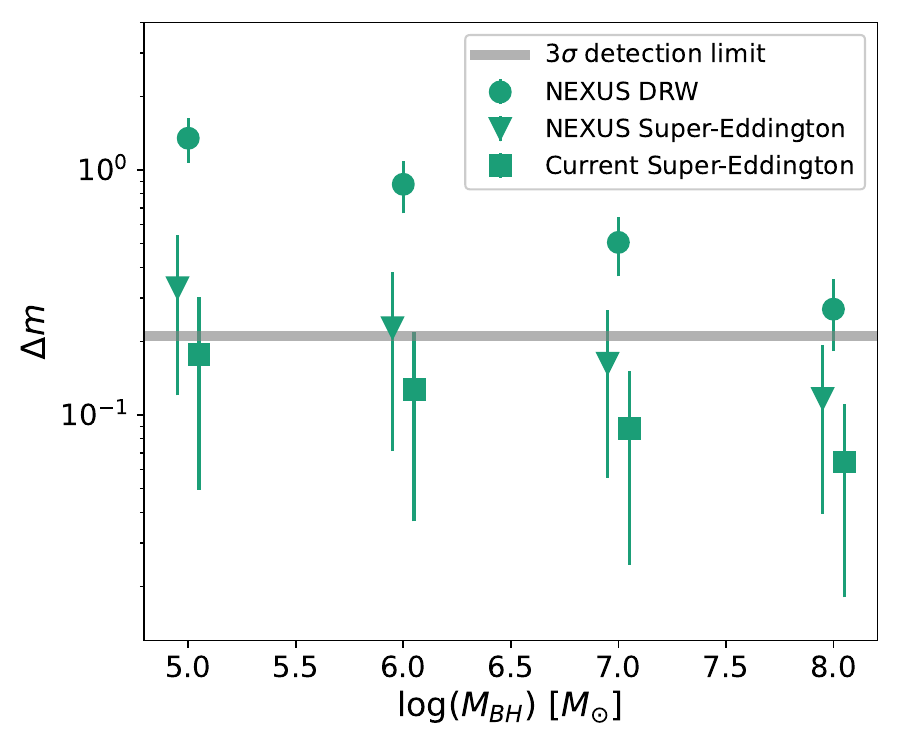}
    \caption{The maximum difference in magnitude, $\Delta m = \max(m) - \min(m)$, between all epochs as a function of mass for 1000 simulated observer-frame F356W-band light curves. Filled circles show the mean $\Delta m$ for mock sub-Eddington DRW model {\sc nexus} light curves. The triangles (squares) show the mean $\Delta m$ we estimate for our super-Eddington model {\sc nexus} (\citetalias{Kokubo:2024}) light curves using the scaling SF$_{\infty} \propto M^{-0.15}$ from Equation \ref{eq:sfinf} for a constant Eddington ratio. Error bars show the standard deviations of $\Delta m$ for the 1000 mock light curves and points have been slightly offset to aid legibility. The solid gray line shows our $\sim 3\sigma$ detection limit for variability. If $\Delta m$ decreases with mass for super-Eddington AGN with a scaling similar to sub-Eddington AGN, we should expect to observe variability for the lowest mass super-Eddington LRDs with {\sc nexus}, but not necessarily with current observations.}
    \label{fig:nexus_opt}
\end{figure}

We find that our super-Eddington AGN model for LRD variability agrees better with the current lack of observed continuum variability in LRDs than sub-Eddington empirical models based on lower redshift AGN. In addition, we find that even with ongoing higher cadence, longer baseline observations, such as the {\sc nexus} survey, we will still not observe significant continuum variability for a majority of LRDs if they are super-Eddington AGN, because we do not expect super-Eddington AGN to vary significantly on rest-frame monthly timescales. However, because our extrapolated model PSD for super-Eddington optical variability is $\rm{PSD}\propto \nu^{-3.0}$, we do expect significantly more variability on rest-frame timescales of a few years. The ideal experiment to attempt to detect LRD continuum variability would then require rest-frame multi-year optical observations. 

To study the impact of a longer observation baseline, we simulate 1000 super-Eddington rest-frame optical LRD light curves using the methods in Section \ref{sec:methods:theoretical} and mock observe them twice annually over the 10~year planned JWST lifetime. We show the mean and standard deviation of the maximum change in magnitude for these mock light curves in Figure \ref{fig:summary}. Increasing the baseline of observation to 10~years in the observer-frame gives a $\sim 67\%$ chance of observing variability greater than our $3\sigma$ detection limit out to $z\lesssim9$. Therefore, our models predict that we should be able to observe optical continuum variability towards the end of the planned lifetime of JWST even if LRDs are super-Eddington AGN.

Another alternative would be to try to observe the variability in lensed LRDs, such as the triply lensed LRD A2744-QSO1  \citep{Furtak:2025, ji2025}. Due to lensing-induced time delays between the three images, the observations of A2744-QSO1 span 2.7~years in the rest-frame. Although technically the predictions of our super-Eddington model are in good agreement with the $\Delta m \approx 0.4$~mag measured by \cite{Furtak:2025}, unfortunately this observed change in magnitude is consistent with the uncertainty of the measurement due to the large lensing systematics. 

Due to the inconsistencies between LRDs and lower redshift AGN, others have proposed modifications to standard AGN models. For example, our super-Eddington model agrees well with the black hole star model of \cite{Naidu:2025}, which resembles a super-Eddington AGN with a puffed-up photosphere. Our super-Eddington model light curves could also be relevant to the late stage quasi-star model of \cite{Begelman:2025}, which involves SMBHs accreting from massive envelopes at a super-Eddington rate. Various non-AGN models for LRDs also exist, such as tidal disruption events in collapsing clusters \citep{Bellovary:2025} or inelastic Raman scattering of stellar UV emission \citep{Kokubo:2024}.

There are several caveats to our super-Eddington variability model, which is based on a limited number of computationally expensive radiation MHD simulations. First, in order to simulate the extended optical emission region, we have to infer that the scaling of the variability between the inner and outer disk remains constant and is well represented by the change from $10-200~r_{\rm g}$. As an alternative, we can assume that the variability amplitude remains constant between the UV and optical emitting regions, but the variability timescale scales with the local dynamical timescales, $\propto r^{3/2}$. Using mock optical light curves generated with this method, we find that the maximum change in magnitude decreases even further, on average by roughly two orders of magnitude, compared to our simulations in Section \ref{sec:results}. Therefore, it would be even more unlikely to detect variability with this alternative scaling.

In addition, the super-Eddington radiation MHD simulations that we use are for a $M_{BH}=10^8~M_{\odot}$ SMBH, which is on the more massive side for LRDs. The filled circles in Figure \ref{fig:nexus_opt} show the mean and standard deviation of $\Delta m$ for 1000 sub-Eddington DRW model light curves as a function of SMBH mass. In this model $\Delta m$ decreases with mass because, for a constant accretion rate, SF$_{\infty} \propto M_{BH}^{-0.15}$ (see Equation \ref{eq:sfinf}).

We do not necessarily expect variability to decrease with mass in exactly the same way for our super-Eddington model as it does in the DRW model. However, as an example, we assume $C = C_0M_{BH}^{-0.15}$ and show the mean and standard deviation of $\Delta m(M_{BH})$ for 1000 mock \citetalias{Kokubo:2024} and 1000 mock {\sc nexus} super-Eddington model light curves as a function of SMBH mass, as squares and triangles in Figure \ref{fig:nexus_opt}, respectively. If this scaling is accurate, we could have already observed variability for $M_{BH}=10^5~M_{\odot}$ LRDs even if they are super-Eddington AGN, because 34\% of mock super-Eddington model \citetalias{Kokubo:2024} light curves at this lowest mass have $\Delta m > 3\sigma$. In addition, $\Delta m > 3\sigma$ for 65\% of our mock super-Eddington model {\sc nexus} $M_{BH}=10^5~M_{\odot}$ light curves. Therefore, it may be possible to observe variability with {\sc nexus} for the lowest mass LRDs even if they super-Eddington accretors, although $\Delta m$ is still significantly lower than for our sub-Eddington model light curves.

If our super-Eddington model for LRDs is correct, future observations of broad line variability with {\sc twinkle} would imply that there must be soft X-ray emission driving this variability that we are unable to observe. This emission must then be strong enough to ionize the broad lines but not strong enough to drive continuum variability in the disk. \cite{Secunda:2025} showed that sufficient amounts of high-energy irradiation incident on the UV-optical emitting region of an AGN disk are needed to drive significant variability in UV-optical light curves, in part due to the intrinsic variability driven by turbulent fluctuations in this part of the disk. If the BLR lacks this intrinsic variability, it may be easier for lower levels of X-ray irradiation to drive emission line variability. Future work should also examine whether the geometry of the BLR or clumpiness of super-Eddington driven outflows \citep{Kobayashi:2018} allows larger amounts of X-ray irradiation to reach the BLR while the accretion disk is shielded by a puffed-up photosphere and optically thick outflows. We do not consider hard X-rays as the main driver of broad line variability in the super-Eddington case, because models of super-Eddington AGN suggest they will be hard X-ray weak \citep[e.g.,][]{inayoshi24b}. However, there is evidence that the relative hard X-ray variability is stronger in super-Eddington accretors \citep[e.g.,][]{inayoshi24b}, which could produce larger variations in emission lines.

Our empirical DRW models show that even with only 2 to 4 epochs of observations we should be able to detect variability in sub-Eddington AGN. The high cadence {\sc nexus} survey should then lead to the identification of dozens of non-LRD sub-Eddington high redshift AGN using variability \citep{Shen:2024}. Our models suggest that variability could help distinguish between sub-Eddington and a few times super-Eddington high redshift AGN. However, we caution that changes in variability can also depend on other AGN properties such as X-ray luminosity and SMBH mass. Continued work studying lower redshift super-Eddington AGN can help to better understand how accretion rate impacts variability.

We mainly discuss the lack of continuum variability observed thus far for LRDs. However, continuum variability has been observed for a few LRDs. For example, \cite{Naidu:2025} claim to find a 30\% optical flux variation between two observations for the LRD they base their black hole star model on, although these flux measurements were taken with two different instruments. While more than 300 LRDs examined in \cite{Zhang:2024} did not show significant variability, \cite{Zhang:2024} found significant variability that was correlated between different waveband JWST observations for two LRDs. This rate is actually in good agreement with the results for our super-Eddington model, which also have $\Delta m>3\sigma$ for $\lesssim 1\%$ of our mock \citetalias{Kokubo:2024} light curves. Alternatively, \cite{Naidu:2025} propose that changes in optical flux could be the result of a change in the gas surrounding the LRD. If ongoing surveys fail to observe variability for a majority of LRDs, current and future exceptional cases which do exhibit variability could perhaps provide information on how LRDs transform from gas-enshrouded super-Eddington accretors to sub-Eddington AGN more similar to lower redshift AGN.

\section{Conclusion}
\label{sec:conclude}

A significant amount of JWST observing time continues to be dedicated to understanding the nature of LRDs and determining whether they truly are high redshift AGN. Particularly intriguing is the current lack of detected variability for nearly all LRDs that have multiple epochs of observation, because variability is an unambiguous identifier of AGN. However, we need to understand whether this lack of variability can be explained by a combination of the insufficient time sampling and short rest-frame baseline of current observations or different physical properties of LRDs, such as super-Eddington accretion rates. 

Here we generate mock light curves using both empirical DRW models for AGN variability based on lower redshift, sub-Eddington AGN and models derived from radiation MHD simulations of super-Eddington AGN disks. We find that our simulations based on empirical sub-Eddington models predict a larger amount of variability than has been observed so far at both rest-frame UV and optical wavelengths. The ongoing {\sc nexus} campaign will provide the highest cadence observations of dozens of LRDs to date. The simulated light curves presented in this paper predict that if most of the LRDs observed in the {\sc nexus} survey do not show significant variability, $\Delta m>0.21$~mag, then either LRDs are not AGN, or LRD variability is not well described by traditional DRW models. 

Current observational constraints on LRD variability are in much better agreement with our mock light curves modeled as super-Eddington AGN. In this model, the UV emission from the AGN is absorbed or scattered, and we find that the optical emission only weakly varies on the relevant timescales. As a result, we do not detect significant variability for 99\% of mock super-Eddington light curves with current observing baselines and cadences and 88\% of mock super-Eddington {\sc nexus} light curves. These simulated light curves are only a few times Eddington, so it is only necessary for LRDs to be moderately super-Eddington to sufficiently suppress variability even if we continue to not observe variability from LRDs with the {\sc nexus} campaign.

The most promising way to use variability to help confirm if LRDs are AGN may be to look for variability in broad line emission, because the observation of broad lines from numerous LRDs suggests that ionizing radiation from the inner rapidly time-varying disk must reach the BLR. We show that if this ionizing radiation comes from the soft X-ray emitting region in the inner disk, the ongoing JWST campaign {\sc twinkle} should observe broad line variability for at least 4, and most likely 8 or 9, of the 9 LRDs that it will observe. On the other hand, the ionizing radiation from the UV emitting region at $50-200~r_{\rm g}$ will not drive enough variability in the broad lines to be observed with {\sc twinkle}. Therefore studying broad line variability may allow us to study X-ray emission from LRDs we are unable to directly observe.

 Overall, we find that a super-Eddington accretion model can easily account for the lack of observed variability in LRDs. Super-Eddington accretion can also help solve several other LRD mysteries. For example, super-Eddington accretion can help relax the need for very heavy black hole seeds \citep{kokorev23,furtak24b,wang24a,Lupi:2024}, and perhaps explain the lack of observed X-ray emission \citep[e.g.,][]{ananna24,yue24,kocevski24} and high-ionization emission lines \citep{lambrides24,madau25}. In addition, super-Eddington accretion naturally leads to optically thick outflows and a puffed up photosphere \citep[e.g.,][]{Shakura1973,King:2003,Jiang:2024,Jiang:2025}, leading to gas absorption that could help explain the odd SEDs of these objects \citep{inayoshi24a,Naidu:2025,Maiolino:2025}. In fact, \cite{Liu:2025} find the Balmer breaks and reddening of LRD SEDs are natural consequences of super-Eddington accreting AGN \citep[see also,][]{Kido:2025}. Uncovering whether LRDs are super-Eddington accreting AGN will have a large impact on understanding black hole seeding and growth, and increase our understanding of galaxy evolution within the first 1 billion years after the Big Bang.

\begin{acknowledgements}
    The authors would like to thank the reviewer for their helpful comments and Dr. Rohan Naidu for helpful information on {\sc twinkle}. The Center for Computational Astrophysics at the Flatiron Institute is supported by the Simons Foundation. AZ acknowledges support by Grant No. 2020750 from the United States-Israel Binational Science Foundation (BSF) and Grant No. 2109066 from the United States National Science Foundation (NSF); and by the Israel Science Foundation Grant No. 864/23.
\end{acknowledgements}

\appendix

\section{Example Mock Light Curves}
\label{appendix}

\begin{figure*}
    \centering
    \includegraphics[width=\textwidth]{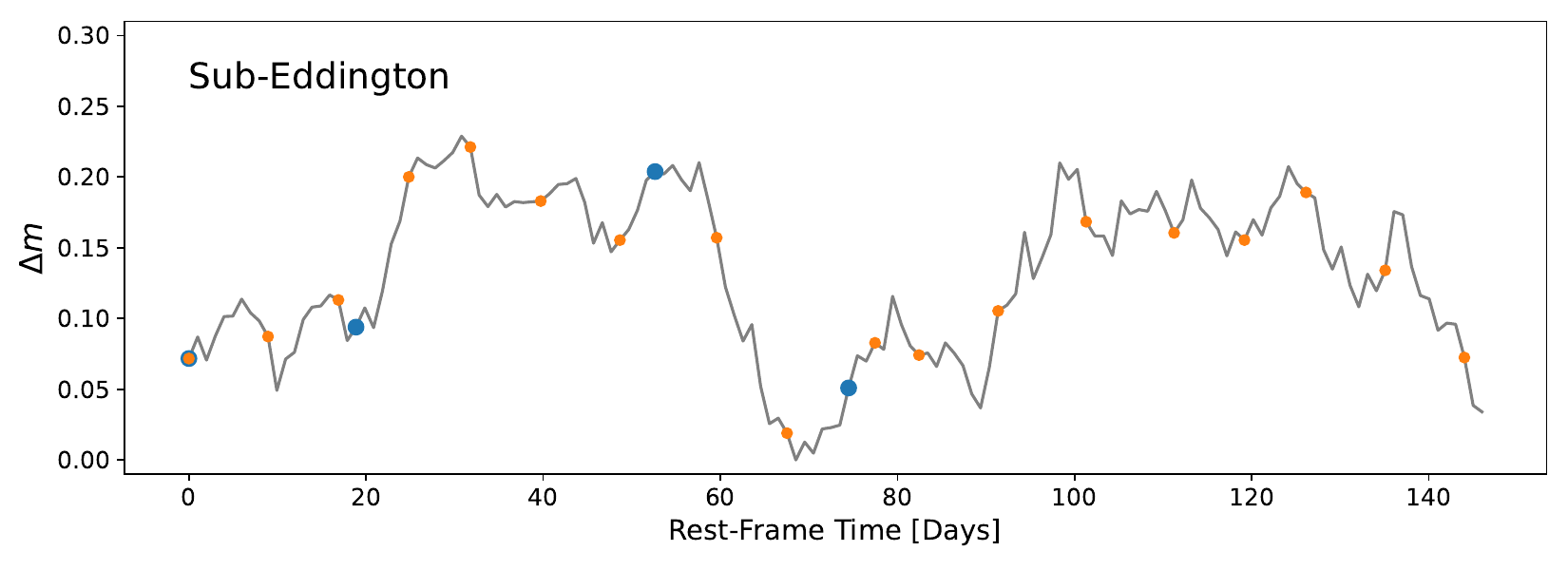} \\
    \includegraphics[width=\textwidth]{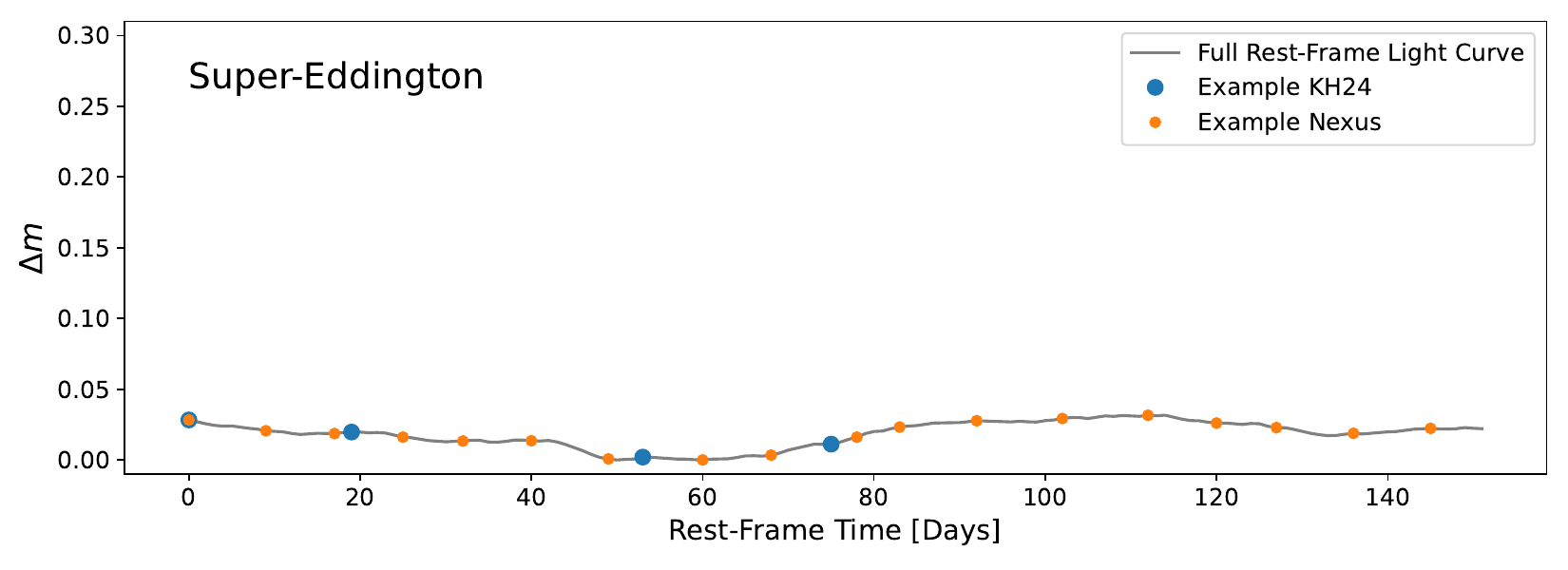}
    \caption{Example rest-frame optical light curves for a $10^8~M_{\odot}$ AGN at $z=6$ generated using our sub-Eddington DRW model (top panel) and our super-Eddington model (bottom panel). We also show the observed times for an example \citetalias{Kokubo:2024} and {\sc nexus} light curve starting at $t=0$ for this mock light curve as the blue and orange dots, respectively. The sub-Eddington light curve has significantly more variability over this rest-frame timescale.}
    \label{fig:exlcs}
\end{figure*}

In this appendix we provide examples of the sub-Eddington and super-Eddington optical and broad emission line light curves discussed in this paper. Figure \ref{fig:exlcs} shows an example rest-frame optical light curve for a $10^8~M_{\odot}$ AGN at $z=6$ modeled using our empirical DRW model for sub-Eddington AGN (top panel) and our model based on the AGN simulations in \cite{Jiang:2025} for super-Eddington AGN (bottom panel). We show the PSD we use to model the sub-Eddington light curve as the dashed line in Figure \ref{fig:super_psd} and the PSD we use to model the super-Eddington light curve as the solid gray line in Figure \ref{fig:super_psd}. In Figure \ref{fig:exlcs} we also show the times at which we mock observe these light curves based on the observation times for an example LRD from \citetalias{Kokubo:2024} (in blue) and the anticipated {\sc nexus} cadence (in orange).

The DRW model sub-Eddington AGN light curve has more variability than the super-Eddington model light curve on all timescales shorter than the {\sc nexus} observing baseline. As a result, {\sc nexus} would be able to observe $\Delta m=0.20$ for the sub-Eddington light curve while $\Delta m=0.030$ for the super-Eddington light curve. The observing cadence of {\sc nexus} is high enough that these $\Delta m$ are similar to the total change in magnitude over the observing baseline. However, as we discuss in Section \ref{sec:discuss}, the super-Eddington light curves have more variability on multi-year long timescales in the rest-frame.  Therefore, the baseline of {\sc nexus}, not the cadence, is the limiting factor in observing more variability in super-Eddington light curves.

The shorter baseline and fewer observations for current multi-epoch LRD observations, like the example cadence for \citetalias{Kokubo:2024} we show in Figure \ref{fig:exlcs}, do make it more difficult to observe a larger change in magnitude. We would observe $\Delta m = 0.15$ and $\Delta m = 0.026$ for this \citetalias{Kokubo:2024} observing cadence for these sub- and super-Eddington light curves, respectively. However, it is still the PSD of the light curve that determines how much variability is observed. These light curves help illustrate that current and ongoing observations provide an accurate approximation for the amount of variability in the full light curve over the observing baseline making it much easier to observe variability in sub-Eddington AGN light curves.

\begin{figure*}
    \centering
    \includegraphics[width=\textwidth]{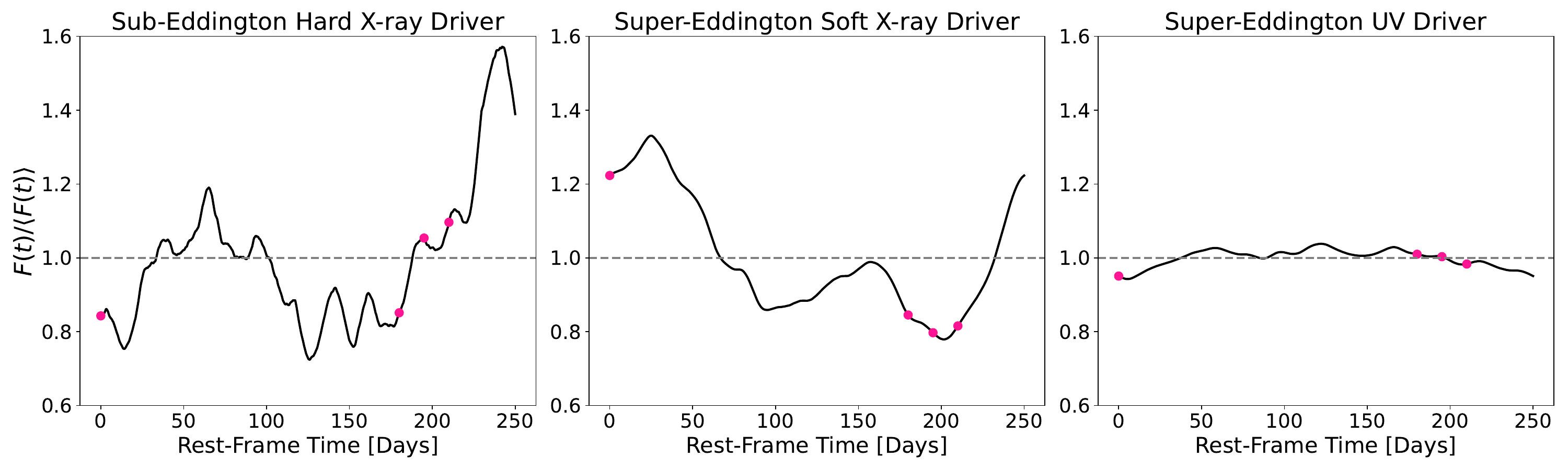}
    \caption{The fractional change in flux as a function of rest-frame time, $F(t)/\langle F(t) \rangle$, for three example light curves for our three different models for reprocessed broad line emission for a $10^8~M_{\odot}$, $z=5.3$ AGN. The difference between the three models is the underlying driving light curve which from left to right is a sub-Eddington model hard X-ray driver, super-Eddington model soft X-ray driver, and super-Eddington model UV driver. The pink dots are the anticipated observation times of {\sc twinkle}. The hard and soft X-ray light curves have sufficient variability to be observed by {\sc twinkle} while the UV light curve does not.}
    \label{fig:exblr}
\end{figure*}

Figure \ref{fig:exblr} shows three example light curves for our three models for reprocessed broad line emission from a $10^8~M_{\odot}$ AGN at $z=5.3$. We scale the flux as a function of time, $F(t)$, by the mean flux of the light curve, $F(t)/\langle F(t) \rangle$, to see the fractional change in flux more easily. The left panel shows the broad emission line light curve for a sub-Eddington hard X-ray driver, the middle panel shows the broad emission line light curve for a super-Eddington soft X-ray driver, and the right panel shows the broad emission line light curve for a super-Eddington UV driver (see Section \ref{sec:methods:emission} for details). The pink points show the mock observing times of {\sc twinkle}.

The broad emission line light curve with a hard X-ray driving light curve has a greater fractional change in flux than the light curve with a soft X-ray driver, but both broad emission line light curves have enough fluctuations $>15\%$ that {\sc twinkle} should expect to observe variability in both. On the other hand, our emission line light curve modeled with a UV driving light curve only fluctuates 10\% over the entire observing baseline of {\sc twinkle}. Due to the sparsity of observations, {\sc twinkle} will only observe a 6\% fluctuation, which is far below the threshold for a significant detection. These example broad emission line light curves illustrate the conclusions of Section \ref{sec:results:future:line}, that if we observe variability in emission lines, then the underlying driving light curve must be the hard or soft X-ray light curve.

\bibliography{agn_lag.bib}
\end{CJK*}
\end{document}